\documentclass[12pt]{article}
\usepackage[]{epsfig}
\usepackage[centertags]{amsmath}
\usepackage{amsfonts}
\usepackage{amssymb}
\usepackage[english]{babel}
\usepackage{amsmath, amsthm, slashed,url}
\usepackage{color}
\begin{document}

%%%%%%%%%%%%%%%%%%%%%%%%%%%%%%%%%%%%%%%%%%%%%%%%%%%%%%%%%%%%%%%%%%%
\title{\bf Landau levels for graphene layers in noncommutative plane}

\author{ 
Lucas~Sourrouille
\\
{\normalsize\it  IFLP-CONICET
}\\ {\normalsize\it Diagonal 113 y 64 (1900), La Plata, Buenos Aires, Argentina}
\\
{\footnotesize lsourrouille@iflp.unlp.edu.ar} } \maketitle

\abstract{Starting from the zero modes of the single and bilayer graphene Hamiltonians we develop a mechanism to construct the 
eigenstates and eigenenergies for Landau levels in  noncommutative plane. General formulas for the spectrum of energies are deduced, for both cases, single and bilayer 
graphene. In 
both cases we find that the effect to introduce noncommutative coordinates is a shift in the energy spectrum with respect to result obtained in commutative space. 
}

\vspace{0.3cm}
{\bf Keywords}: Quantum Mechanics; Noncommutative field theory; Mathematical methods 
in Physics; Graphene; Landau Levels; Zero Energy Modes

%%%%%%%%%%%%%%%%%%%%%%%%%%%%%%%%%%%%%%%%

%%%%%%%%%%%%%%%%%%%%%%%%%%%%%%%%%%%%%%%%
\vspace{1cm}
\section{Introduction}
The study of a single electron in two dimensions interacting with a magnetic field was analyzed manly by Landau, Darwin and Fock 
\cite{L}-\cite{F}, showing that the electron kinetic energy is quantized. Such, quantization, usually namely Landau levels is 
important in many problems in quantum physics, such as, the quantum Hall problem \cite{3, 4, 5.1, 5.3, 
5.5, 5.6, 5.7}. 
\\
In the last years, experimental realization of monolayer graphene films \cite{1.1,2.1,3.1}, opened a new area in condensed matter 
physics. In particular, these works revealed that graphene electrons behave like massless Dirac fermions, that is like 
relativistic particles. Thus, the relativistic behavior of electron in graphene open the possibilities
for testing relativistic phenomena in condensed matter, some of which are unobservable in high-energy physics. One of the most 
interesting phenomena in graphene is the anomalous Landau-Hall effect  \cite{3.2} and minimum quantum conductivity in the 
limit of vanishing concentration of charge carriers \cite{1.1}. Other, relativistic phenomena, such as Klein-Gordon paradox, 
could 
be tested in graphene \cite{4.1,4.2,4.3}. In this context, Landau levels in graphene also presents new features \cite{6}.
\\
On the other hand, the study of field theories in noncommutative space, also, has received much attention in the last few years 
\cite{rew1}-\cite{rew3}.  
The connection between these theories and string theory was first considered by Connes, Douglas and Schwartz \cite{CDS}, who 
observed that 
noncommutative geometry arise as a possible scenario for certain low energy of string theory and M-theory. Afterwards Seiberg and 
Witten showed that the low energy dynamics of string theory can be described in terms of the noncommutative Yang-Mills theory 
\cite{witten1}. Since then many papers appeared covering diverse applications of noncommutative theory in physical problems.
\\
In the present Letter, we deal with a scenario in which the coordinates of the plane are noncommutative. A few articles explore the problem of graphene when the space or phase-space becomes noncommutative \cite{ref1,ref2,ref3}. This articles use the Seiberg-Witten map as tool, however they did not be able to give an expression of the energy spectrum in noncommutative space. In Section \ref{3v} we discuss general properties of the Dirac-Weyl equation for Landau Levels in the symmetric gauge, emphasizing the construction of zero energy mode.
Section \ref{4v}, is addressed to develop a method that allow us to construct the eigenenergies and the eigenstates for 
Landau levels in the case of single layer graphene in noncommutative plane. We will find that general formula for the spectrum of 
the energy is given by 
\begin{eqnarray}
E = \pm 
\sqrt{3NB} \;, \;\;\;
\;\;\; N= 
0,1,2,...
\label{}
\end{eqnarray}
which implies that the spectrum is shifted by $\pm (\sqrt{3} -\sqrt{2})\sqrt{NB}$  with respect of the spectrum of the 
commutative case. 
Finally, we analyze, in Section \ref{5v}, the application of the method developed in the previous section, to the case of bilayer 
graphene Hamiltonian. We will be able to deduce a general formula for the spectrum of the Landau energy. Particularly, we will 
find that the spectrum in noncommutative space is dictated by the formula,
\begin{eqnarray}
E = \pm\sqrt{9 B^2 N(N-1)},\;\;\;\; N= 0,1,2,3...
\label{}
\end{eqnarray}
As in the case of single layer, we will note that the levels of enegry are shifted with respect of the result of the commutative 
case. Specifically, we will show that the shift is $\pm(\sqrt{9}-\sqrt{4})\sqrt{B^2 N(N-1)}$.

\section{The commutative massless Dirac fermions in a magnetic field }
\label{3v}

One of the interesting aspects of graphene is that a Dirac electron moves with an effective Fermi
velocity $v_F = c/ 300$, where c is the velocity of light,
and behaves as a massless quasi-particle. The effective
Hamiltonian, in $(2+1)$-dimensions, around a Dirac point for a Dirac electron has the form [2]
\begin{equation}
H= v_F\sigma^i p_i = v_F(\sigma^1 p_1 +  \sigma^2 p_2)\;,
\label{}
\end{equation}
Here, the $\sigma^i$ $(i =1,2)$ 
are 2$\times$2 Pauli matrices, i.e.
\begin{eqnarray}
\sigma^1 =\left( \begin{array}{cc}
0 & 1 \\
1 & 0 \end{array} \right)
\,,
\;\;\;\;\;\
\sigma^2 =\left( \begin{array}{cc}
0 & -i \\
i & 0 \end{array} \right)
\end{eqnarray}
and $p_i =-i\partial_i$ is the two-dimensional momentum operator. The massless Dirac-Weyl equation in $(2+1)$ dimensions is
\begin{equation}
v_F\sigma^i p_i \Phi (x, y, t) = i\hbar\partial_t \Phi (x, y, t)
\label{eq1}
\end{equation}
(For simplicity we will choose in this paper  $v_F=\hbar=1$).
Here, $\Phi (x, y, t)$ is the two-component spinor
\begin{equation}
\Phi=(\phi_a,\phi_b)^T
\label{}
\end{equation}
where $\phi_a$ and $\phi_b$ represent the envelope functions associated with the probability amplitudes. Since, we are interested 
in stationary states, it is natural to propose a 
solution of the form
\begin{eqnarray}
\Phi (x, y, t) = e^{-iEt} \Psi (x, y)\;,
\label{}
\end{eqnarray}
then, the time-independent Dirac-Weyl equation is
\begin{equation}
\sigma^i p_i \Psi (x, y) = E \Psi (x, y)
\label{1dw}
\end{equation}
In the presences of a perpendicular magnetic field to the $(x, y)$-plane, we replace the momentum operator $p_i$ by the covariant 
derivative, defined as $D_{i}= -i\partial_{i} +A_{i}$ $(i =1,2)$, where $A_{i}$ are components of the vector potential,
\begin{equation}
B=\partial_x A_y -\partial_y A_x
\label{mag}
\end{equation}
Thus, the equation (\ref{1dw}) becomes, 
\begin{equation}
\sigma^i D_i \Psi (x, y) = E \Psi (x, y)
\label{2dw}
\end{equation}
By developing this equation to get,
\begin{equation}
\left( \begin{array}{cc}
0 & D_1 -iD_2 \\
D_1 +iD_2 & 0 \end{array} \right) \left( \begin{array}{c}
\psi_a\\
\psi_b \end{array} \right) = E \left( \begin{array}{c}
\psi_a\\
\psi_b \end{array} \right) 
\label{3dw}
\end{equation}
where $\psi_a$ and $\psi_b$ are the components of the spinor $\Psi$ (i.e. $\Psi=(\psi_a,\psi_b)^T$). From this equation we can
write the two coupled equations for the components $\psi_a$ and $\psi_b$
\begin{eqnarray}
D_1 \psi_b -iD_2 \psi_b = E \psi_a
\label{eqm1}
\end{eqnarray}
\begin{eqnarray}
D_1 \psi_a +iD_2 \psi_a = E \psi_b
\label{eqm2}
\end{eqnarray}
Here, we are interested to find the the eigenvalues and eigenstates corresponding to Eq.(\ref{2dw}). In order to find these one 
needs to specify a gauge for the vector potential. Here, we will use the symmetric gauge,
\begin{eqnarray}
{\bf A} = \frac{1}{2} (-By , Bx, 0)
\label{vec}
\end{eqnarray}
Then, equations (\ref{eqm1}) and (\ref{eqm2}) becomes, 
\begin{eqnarray}
\Big[ \partial_z - z\frac{B}{2} \Big] \psi_b = E \psi_a
\label{eq1.}
\end{eqnarray}
\begin{eqnarray}
\Big[ \partial_{z^{\dagger}}- z^{\dagger}\frac{B}{2} \Big] \psi_a = E \psi_b
\label{eq2.}
\end{eqnarray}
where $z= ix + y$, $z^{\dagger}= -ix + y$, $\partial_z = -i\partial_x -\partial_y$ and $\partial_{z^{\dagger}} = -i\partial_x +\partial_y$. The simplest solution of the equations (\ref{eq1.}) and (\ref{eq2.}) are the 
zero energy modes, that is the solutions for zero energy. Let us start, by constructing the solutions for zero energy. 
For this purpose we assume that the vector 
potential is divergenceless, which is clearly 
satisfied by (\ref{vec}). Then, one can introduce a scalar potential $\lambda(x, y)= \frac{B}{4}(x^2 + y^2)$ such that, 
\begin{eqnarray}
A_x = -\partial_y \lambda
\,,
\;\;\;\;\;\
A_y = \partial_x \lambda
\label{gau}
\end{eqnarray}
so that, the magnetic field reads,
\begin{eqnarray}
B = \partial_x^2 \lambda + \partial_y^2 \lambda
\label{div}
\end{eqnarray}
In view of (\ref{gau}), it is not difficult to find the solutions of the equations (\ref{eq1.}) and (\ref{eq2.}) for zero energy 
case. 

\begin{eqnarray}
\psi_{a,b} = f_{a,b} e^{\frac{\gamma B}{4}(x^2 + y^2)}
\label{sol}
\end{eqnarray}
where$f_a$ and $f_b$ are complex conjugated analytic entire functions of $z$. 
Here, $\gamma =1$ and $-1$ for $\psi_a$ and $\psi_b$ respectively. Since, the function $f(z)$ can not go to zero in all 
directions at infinity, $\psi_{a,b}$ is normalizable only if we assume that $\gamma B < 0$. In other words, the zero energy 
modes 
exist only for one spin direction, which depend on the sign of the magnetic field.
\\
Finally it is necessary to mark that the functions $f_{a,b}$ are  
polynomials of the form
\begin{eqnarray}
f_{a} =\tilde{a}_i (z^\dagger)^i \;, \;\;\;
\;\;\; i= 
0,1,2,...
\end{eqnarray}
\begin{eqnarray}
f_{b} = a_i z^i \;, \;\;\;
\;\;\; i= 
0,1,2,...
\end{eqnarray}
being $\tilde{a}_i$ and $a_i$ real numbers. 

\section{The construction of the eigenvalues and eigenstates for a negative magnetic field in noncommutative plane: the case of 
single layer graphene}
\label{4v}
Let us call $x_\mu$ , $\mu = 1, 2, ...d$ the coordinates of d-
dimensional space-time. Given $f(x)$ and $g(x)$, two ordinary
functions in $R^d$ , their Moyal product is defined as \cite{5},\cite{rew2}
\begin{eqnarray}
(f* g)(x) &=&  \exp(\frac{i}{2}\theta_{\mu\nu} \partial_x^\mu \partial_y^\nu) f(x) g(x)|_{y=x}
\nonumber \\
&=&
f(x)g(x) + \frac{i}{2} \theta_{\mu\nu} \partial^\mu f(x) \partial^\nu g(x) 
\nonumber \\
&-& \frac{1}{8}\theta_{\mu\alpha}\theta_{\nu\beta} \partial^\mu \partial^\alpha f(x)\partial^\nu \partial^\beta g(x)+...
\label{estrella}
\end{eqnarray}
with $\theta_{\mu\alpha}$ a constant antisymmetric matrix. One can easily see that (\ref{estrella}) defines a noncommutative but associative product,
\begin{eqnarray}
f(x)* (g(x)*h(x))= (f(x)*g(x)) *h(x)
\end{eqnarray}
Under certain conditions, integration over $R^d$ of Moyal products has all the properties of the the trace
\begin{eqnarray}
\int dx^d f(x)* g(x) = \int dx^d g(x)* f(x) =  \int dx^d f(x) g(x)
\end{eqnarray}
One has also in this case cyclic property of the star product,
\begin{eqnarray}
\int dx^d f(x)* g(x)*h(x) =  \int dx^d h(x)*f(x)*g(x)
\end{eqnarray}
Finally, Leibnitz rule holds
\begin{eqnarray}
\partial_\mu (f(x)* g(x))  =  \partial_\mu f(x) *g(x) + f(x)*  \partial_\mu g(x)
\end{eqnarray}
The $*$-commutator, denoted with [ , ],
\begin{eqnarray}
[f(x), g(x)]  =  f(x) *g(x) - g(x)* f(x)
\end{eqnarray}
is usually called a Moyal bracket. If one considers the case
in which $f(x)$ and $g(x)$ correspond to space-time coordinates $x_\mu$
and $x _\nu$ , one has, from (\ref{estrella}), \cite{rew2}
\begin{eqnarray}
[x^\mu, x^\nu]  =  i \theta^{\mu\nu}
\label{co2}
\end{eqnarray}
This justifies the terminology “noncommutative space-time”
although in the Moyal product approach to noncommutative
field theories one takes space as the ordinary one and it is through the star multiplication of fields that noncommutativity
enters into play.
\\
In the special case of two-dimensional space $R^2$ we have coordinates $x^1$ , $x^2$.
In this case we can write, 
\begin{eqnarray}
\theta^{\mu\nu} = \theta \epsilon^{\mu\nu}
\end{eqnarray}
where $\theta$ a constant and $\epsilon^{12}= -\epsilon^{21} = 1$.
Since, we will be interested in $d=2+1$ noncommutative space we can choose $\theta = \frac{-1}{B}$, so that the commutator (\ref{co2}) between $x^1$ and $x^2$ reduce to
\begin{equation}
[x^1,x^2]= \frac{-i }{B},\qquad [x_i,t]=0\;,
\label{1} 
\end{equation}
where $x^1 = x$, $x^2 = y$. 
We, also, assume that spatial coordinates commute with  time,
\begin{equation}
[x_i,t]=0\;\;\;\;, i= 1,2
\label{} 
\end{equation}
In particular we will work with the $z$ and $z^\dagger$ variables defined in the previous section. Therefore, if we use the commutator (\ref{1}) between $x^1$ and $x^2$, we arribe to the following commutator,   
\begin{eqnarray}
[z^\dagger , z] = -\frac{2}{B}
\label{con1}
\end{eqnarray}
In addition, we can calculate, easily, the commutator between the covariant momentums in noncommutative space. This covariant momentums are the covariant derivatives, which was defined above as $D_{i}= -i\partial_{i} +A_{i}$ $(i =1,2)$, so that we have, 
\begin{eqnarray}
[D_x , D_y] = -i \Big( \partial_x A_y -\partial_y A_x + i [A_x, A_y] \Big) = -i B_{NC}
\label{con12}
\end{eqnarray}
where $B_{NC}=  \partial_x A_y -\partial_y A_x + i [A_x, A_y]$ is the expression for the magnetic field in the noncommutative space. It is interested to note that if we compare this expression with the commutative one, the noncommutative magnetic filed incorporate an additional term due to the noncommutative of space.  
\\ 
In order to be  more practical in the notation it is convenient to rewrite the equations (\ref{eq1.}) and (\ref{eq2.}) as 

\begin{eqnarray}
a^\dagger \psi_b = E \psi_a
\label{eq1.1}
\end{eqnarray}
\begin{eqnarray}
a \psi_a = E \psi_b
\label{eq2.1}
\end{eqnarray}
where $a^\dagger = \partial_z - z\frac{B}{2}$ and $a=  \partial_{z^{\dagger}}- z^{\dagger}\frac{B}{2}$. We will see that under a negative magnetic field $a^\dagger$ behaves like a creation operator whereas $a$ acts as destruction operator. 
\\
Here,
we are interested in the construction of  eigenstates and eigenvalues associated to energies different form zero.
In order to illustrate 
our mechanism, we start 
by considering the simplest zero mode solution of the Hamiltonian equations (\ref{eq1.1}) and (\ref{eq2.1}) for a negative magnetic field,
\begin{eqnarray}
\Psi_{0,0} =\left( \begin{array}{c}
\tilde{a}_0 e^{\lambda} \\
0  \end{array} \right)
\label{q+}
\end{eqnarray}
The first subindex denote the number of independent state with zero energy and the second index denote level of 
quantized energy. 
In order to construct an eigenstate associated to first excited level of energy, we can take the operator $a^\dagger$ and apply it to $\psi_a = \tilde{a}_0 e^{\lambda}$, so that,
\begin{eqnarray}
a^\dagger \psi_a = \tilde{a}_0[-i\partial_x  \lambda -\partial_y \lambda]e^\lambda 
-z\frac{B}{2}\tilde{a}_0 e^{\lambda} = -z B \tilde{a}_0 e^{\lambda}
\label{27}
\end{eqnarray}
Thus, we want able to create a new state $-z B \tilde{a}_0 e^{\lambda}$, so that the operator $a^\dagger$ acts on $\psi_a$  as a creation operator. 
Now, applying $a$ upon the state $-z B \tilde{a}_0 e^{\lambda}$ we obtain,
\begin{eqnarray}
a (-z B \tilde{a}_0 e^{\lambda}) &=&  
-2B \tilde{a}_0 e^{\lambda} - \frac{B^2}{2}\tilde{a}_0 z z^\dagger e^{\lambda} + \frac{B^2}{2}\tilde{a}_0 z^\dagger z e^{\lambda} 
\nonumber \\
&=& -2B \tilde{a}_0 e^{\lambda} + \frac{B^2}{2}\tilde{a}_0 [z^\dagger , z] e^{\lambda} 
\label{1.2.}
\end{eqnarray}
By using the commutator (\ref{con1}) we finally obtain,
\begin{eqnarray}
a (-z B \tilde{a}_0 e^{\lambda}) =  
 -3B \tilde{a}_0 e^{\lambda} = -3B\psi_a
 \label{29}
\end{eqnarray}
which show that the operator $a$ acts like an annihilation operator on the state $-z B \tilde{a}_0 e^{\lambda}$. 
Now, we can rewrite the equation (\ref{27}) as 
\begin{eqnarray}
a^\dagger \chi_b = \frac{\pm\sqrt{-3B}}{\pm\sqrt{-3B}}\Big(-z B \tilde{a}_0 e^{\lambda}\Big) = (\pm\sqrt{-3B})  \chi_a
\label{}
\end{eqnarray}
where we have renamed $\psi_a$ as $\chi_b$ and $\frac{-z B \tilde{a}_0 e^{\lambda}}{\pm\sqrt{-3B}}$ as $\chi_a$. Therefore, in view of (\ref{29})  we have,
\begin{eqnarray}
a (\frac{-z B \tilde{a}_0 e^{\lambda}}{\pm\sqrt{-3B}}) =  
 \pm\sqrt{-3B} \tilde{a}_0 e^{\lambda} = \pm\sqrt{-3B} \chi_b
 \label{31}
\end{eqnarray}
Thus, we arrive to the following pair of equations,
\begin{eqnarray}
a^\dagger \chi_b  = (\pm\sqrt{-3B})  \chi_a
\label{}
\end{eqnarray}
\begin{eqnarray}
a \chi_a =  
 \pm\sqrt{-3B} \chi_b
 \label{}
\end{eqnarray}
which shows that 
\begin{eqnarray}
X_{0,1} =\left( \begin{array}{c}
\frac{-z B \tilde{a}_0 e^{\lambda}}{\pm 
\sqrt{-3B}} \\
\tilde{a}_0 e^{\lambda}  \end{array} \right)
\label{q+-}
\end{eqnarray}
is an eigenstate of the Dirac-Weyl Hamiltonian corresponding to the first existed level of energy, that is the level of energy $\pm\sqrt{-3B}$. 
So, we may be able to construct a mechanisms to obtain the energy levels and the eigenstates of the Dirac-Wely equation.  
From (\ref{1.2.}), it is interesting 
to 
note that in the commutative case $[z^\dagger , z]= 0$ and, then, we can deduce easily that the first energy existed level is $E_{c} = \pm \sqrt{-2B}$. 
This is
the well know result for the first Landau level of energy for commutative Dirac fermions (see reference \cite{6}).
\\
In general, we can repeat the same process starting from arbitrary zero mode for a negative magnetic field configuration. In that 
case we take the $i$-degenerate zero energy state, that is,
\begin{eqnarray}
\Psi_{i,0} =\left( \begin{array}{c}
\tilde{a}_i (z^\dagger)^i e^{\lambda} \\
0  \end{array} \right)
\label{+..}
\end{eqnarray}
Applying the operator $a^\dagger$ on the state $\tilde{a}_i (z^\dagger)^i e^{\lambda}$ we obtain
\begin{eqnarray}
a^\dagger(\tilde{a}_i (z^\dagger)^i e^{\lambda})= -2i\tilde{a}_i (z^\dagger)^{i-1} 
e^{\lambda} - \frac{B}{2} \tilde{a}_i[(z^\dagger)^i z + z (z^\dagger)^i] e^{\lambda}
\label{res1.}
\end{eqnarray}
Taking this result and applying the operator $a$, we have, after some algebra, the following expression,
\begin{eqnarray}
&&a\Big(-2i\tilde{a}_i (z^\dagger)^{i-1} 
e^{\lambda} - \frac{B}{2} \tilde{a}_i[(z^\dagger)^i z + z (z^\dagger)^i] e^{\lambda}\Big) \label{res2}
\nonumber \\
&& = - 2\tilde{a}_i B (z^\dagger)^i e^{\lambda} + \tilde{a}_i \Big(\frac{B}{2}\Big)^2 \Big( [z^\dagger, [(z^\dagger)^i z + z(z^\dagger)^i] \Big)\label{res2}
\end{eqnarray}
where the commutator in the last expression may be developed to give,
\begin{eqnarray}
\Big[z^\dagger, [(z^\dagger)^i z + z(z^\dagger)^i]\Big] = - \Big[z^\dagger, z^\dagger z z^\dagger\Big](z^\dagger)^{i-2} - 
(z^\dagger)^{i-2}\Big[z, (z^\dagger)^3\Big]
\label{}
\end{eqnarray}
The two brackets in the right hand can be calculated explicitly, 
\begin{eqnarray}
\Big[z^\dagger, z^\dagger z z^\dagger\Big] = -\frac{2}{B} (z^\dagger)^2
\nonumber \\[3mm]
\Big[z, (z^\dagger)^3\Big] =  \frac{6}{B}(z^\dagger)^2
\label{}
\end{eqnarray}
so that, 
\begin{eqnarray}
\Big[z^\dagger, [(z^\dagger)^i z + z(z^\dagger)^i]\Big] = -4\frac{(z^\dagger)^i}{B}
\label{}
\end{eqnarray}
Then, we can write the term in right hand of the equality (\ref{res2}),
\begin{eqnarray}
&&a\Big(-2i\tilde{a}_i (z^\dagger)^{i-1} 
e^{\lambda} - \frac{B}{2} \tilde{a}_i[(z^\dagger)^i z + z (z^\dagger)^i] e^{\lambda}\Big) 
\nonumber \\
&&= - 2\tilde{a}_i B (z^\dagger)^i e^{\lambda} - \tilde{a}_i B (z^\dagger)^i e^{\lambda} = -3B ( \tilde{a}_i (z^\dagger)^i 
e^{\lambda})
\label{}
\end{eqnarray} 
which show that $\tilde{a}_i (z^\dagger)^i e^{\lambda}$ is an eigenstate of $aa^\dagger$ with eigenvalue $-3B$. Therefore, we can write
\begin{eqnarray}
a^\dagger \chi_{i,b}  = (\pm\sqrt{-3B})  \chi_{i,a}
\label{}
\end{eqnarray}
\begin{eqnarray}
a \chi_{i,a} =  
 \pm\sqrt{-3B} \chi_{i,b}
 \label{}
\end{eqnarray}
where  
\begin{eqnarray}
\chi_{i,a} = \frac{\tilde{a}_i}{\pm \sqrt{-3B}} \Big[-2i (z^\dagger)^{i-1} 
e^{\lambda} - \frac{B}{2}[(z^\dagger)^i z + z (z^\dagger)^i] e^{\lambda}\Big]
\end{eqnarray}
\begin{eqnarray}
\chi_{i,b} = \tilde{a}_i (z^\dagger)^i e^{\lambda}
\end{eqnarray}
Hence, we have obtained, a general solution for noncommutative plane, one for each $i$, for the eigenstates of the first existed level of energy. In terms of the spinor notation, these states take the form 
\begin{eqnarray}
X_{i,1} =\left( \begin{array}{c}
\frac{\tilde{a}_i}{\pm \sqrt{-3B}} \Big[-2i (z^\dagger)^{i-1} 
e^{\lambda} - \frac{B}{2}[(z^\dagger)^i z + z (z^\dagger)^i] e^{\lambda}\Big] \\
 \tilde{a}_i (z^\dagger)^i e^{\lambda} \end{array} \right)
\label{q+-.}
\end{eqnarray}
We can repeat the mechanism used to obtain the first energy level and its eigenstates, to construct the eigenfunctions of the higher energy levels. Indeed, if we consider the state (\ref{q+-}) and apply the operator $a$ to the upper component, we have,    
\begin{eqnarray}
a^\dagger \Big( \frac{-z B \tilde{a}_0 e^{\lambda}}{\pm 
\sqrt{-3B}} \Big)= \frac{(z B)^2 \tilde{a}_0 e^{\lambda}}{\pm 
\sqrt{-3B}}
\label{m1}
\end{eqnarray}
Following the previous steps we apply the operator $a^{\dagger}$ to the 
result (\ref{m1}), which lead us to  
\begin{eqnarray}
a^{\dagger} \Big(\frac{(z B)^2 \tilde{a}_0 e^{\lambda}}{\pm 
\sqrt{-3B}}\Big) = -4 B \Big(\frac{ -\tilde{a}_0 B z e^{\lambda}}{\pm\sqrt{-3B}}\Big) + \frac{\tilde{a}_0 B^3 [z^2, z^\dagger] 
e^{\lambda}}{2 \pm \sqrt{-B}} - 
\label{44.}
\end{eqnarray}
where the commutator $[z^2, z^\dagger]$ can be calculated easily from (\ref{con1}) to give $[z^2, z^\dagger] = \frac{4}{B}$, therefore,
\begin{eqnarray}
a^{\dagger} \Big(\frac{(z B)^2 \tilde{a}_0 e^{\lambda}}{\pm 
\sqrt{-3B}}\Big) = -6 B \Big(\frac{ -\tilde{a}_0 B z e^{\lambda}}{\pm\sqrt{-3B}}\Big) 
\label{}
\end{eqnarray}
Following the same procedure for the first level of energy we have  the couple of equations 
\begin{eqnarray}
a^\dagger \zeta_b  = (\pm\sqrt{-6B})  \zeta_a
\label{}
\end{eqnarray}
\begin{eqnarray}
a \zeta_a =  
 \pm\sqrt{-6B} \zeta_b
 \label{}
\end{eqnarray}
where, 
\begin{eqnarray}
&&\zeta_a = \frac{(z B)^2 \tilde{a}_0 e^{\lambda}}{\pm 
\sqrt{18B^2}}
\nonumber \\[3mm]
&&\zeta_b = \frac{-z B \tilde{a}_0 e^{\lambda}}{\pm 
\sqrt{-3B}}
\nonumber \\[3mm]
&&E = \pm \sqrt{-6B}
\label{}
\end{eqnarray}
Notice that in the commutative case the last term of (\ref{44.}) is zero, which implies that energy corresponding to second 
excited Landau level is $E_{c} = \pm \sqrt{-4B}$. This result is in accordance to the well know formula
\begin{eqnarray}
E_{c} = \pm \sqrt{-2NB} \;, \;\;\;
\;\;\; N= 
0,1,2,...
\label{62.}
\end{eqnarray}
This formula is the general expression for the spectrum of the eigenenergies for massless Dirac electrons in a uniform magnetic 
\cite{6}. Returning to the noncommutative case, we conclude that the spinor 
\begin{eqnarray}
Z_{0,2} =\left( \begin{array}{c}
\frac{(z B)^2 \tilde{a}_0 e^{\lambda}}{\pm 
\sqrt{18B^2}} \\
\frac{-z B \tilde{a}_0 e^{\lambda}}{\pm 
\sqrt{-3B}}  \end{array} \right)
\label{48.}
\end{eqnarray}
is an eigenstate of the Dirac-Wely equation, being $E = \pm \sqrt{-6B}$ the corresponding eigenvalue associated to the 
second excited energy level.
\\
We can repeat the process by choosing an arbitrary eigenstate corresponding to the first excited energy level. In other words, if 
we want to get the general form of the eigenfunctions corresponding to the second excited energy level we should take the 
state (\ref{q+-.}) and apply the same process that lead us to result (\ref{48.}).
\\
The method may be generalised for higher energy levels. Then, it is not difficult to find an
expression for the spectrum of the eigenenergies, in the 
noncommutative plane,
\begin{eqnarray}
E = \pm 
\sqrt{-3NB} \;, \;\;\;
\;\;\; N= 
0,1,2,...
\label{}
\end{eqnarray}
By comparing this formula with the formula (\ref{62.}),
it is interesting to remark that the effect to become the plane noncommutative is a shift in the energies spectrum.  
\\
To finalize, it is necessary to mention that the mechanism developed in this section may be also apply to the case of positive 
magnetic field. In that case, according to the formulas (\ref{eq1.}) and (\ref{eq2.}) the simplest zero mode is
\begin{eqnarray}
\Psi_{0,0} =\left( \begin{array}{c}
0 \\
a_0 e^{-\lambda}  \end{array} \right)
\label{}
\end{eqnarray}
Then, we can apply to $\psi_b$ the operator $\Big[\partial_{z^\dagger} - z^{\dagger}\frac{B}{2} \Big] = a$ and then 
$\Big[\partial_z - z\frac{B}{2} \Big] = a^\dagger$. Here, notice that the order of the application of the operators is 
inverse to the case of negative magnectic field.
Thus, in this case, the operator $a$ would behave like a creation operator while $ a^\dagger$ would behave like destruction operator. 
For more details, about the application of this mechanism in the case 
of positive magnetic field in commutative space see \cite{my1}.

\section{Application to bilayer graphene}
\label{5v}
Consider now the case of bilayer graphene \cite{6}, \cite{mc}, \cite{gc1}.
Let us start with the simplest Hamiltonian, 
\begin{equation}
H = \left( \begin{array}{cc}
0 & (a^\dagger)^2 \\
a^2 & 0 \end{array} \right) 
\label{3dw3}
\end{equation}
which means intermediate energies.
This description is accurate at the energy scale
larger than a few meV, otherwise a more complicated picture
including trigonal warping takes place. We will restrict ourselves only by the case of not too small doping when the
approximate Hamiltonian (\ref{3dw3}) is valid. 
\\
Then, the eigenvalue equation becomes,
\begin{equation}
\left( \begin{array}{cc}
0 & (a^\dagger)^2 \\
a^2 & 0 \end{array} \right)\left( \begin{array}{c}
\psi_a\\
\psi_b \end{array} \right) = E \left( \begin{array}{c}
\psi_a\\
\psi_b \end{array} \right)
\label{3d12}
\end{equation}
Again, we choose the symmetric gauge (\ref{vec}), so that,
\begin{eqnarray}
(a^\dagger)^2 \psi_b = E \psi_a
\label{bi1}
\end{eqnarray}
\begin{eqnarray}
a^2 \psi_a = E \psi_b
\label{bi2}
\end{eqnarray}
First, we analyze the zero energy modes of these equations.  The solution of these equations, for zero energy, can be found easily 
from the 
development done in Section \ref{3v}. Indeed, we can check that the sets 
\begin{eqnarray}
\left( \begin{array}{c}
 \tilde{a}_i (z^\dagger)^i e^{\lambda}\\
0  \end{array} \right),\;\;\;\;\;\;
\left( \begin{array}{c}
 z\tilde{a}_i (z^\dagger)^i e^{\lambda}\\
0  \end{array} \right)
\label{fb1}
\end{eqnarray}
and  
\begin{eqnarray}
\left( \begin{array}{c}
0\\
a_i z^i e^{-\lambda}  \end{array} \right),\;\;\;\;\;\;
\left( \begin{array}{c}
0\\
z^\dagger  a_i z^i e^{-\lambda}  \end{array} \right)
\label{fb2}
\end{eqnarray}
are solutions of the equations (\ref{bi1}) and (\ref{bi2}) for zero energy. Notice that the zero mode for bilayer are twice as 
great as for the case of a single layer. With this ideas in mind we can construct the spectrum of energies and their 
eigenfunctions, in noncommutative plane, in a similar way to the case of single layer. Continuing with the same scheme as in the 
previous section, we consider the case in which the magnetic field is negative. Thus, we start by considering 
the two simplest solutions given by the equation (\ref{fb1}). In other words, we should take the two following zero modes,
\begin{equation}
\left( \begin{array}{c}
\tilde{a}_0 e^{\lambda}\\
0 \end{array} \right) \;, \;\;\;
\;\;\;
\left( \begin{array}{c}
z \tilde{a}_0 e^{\lambda}\\
0 \end{array} \right)
\label{zerobi}
\end{equation}
Now, we can start, from these zero modes, to construct the eigenfunctions and eigenenergies corresponding to excited levels. 
Following the process developed in Section \ref{4v}, we apply the operator $(a^\dagger)^2$ to the state $\psi_{a} = \tilde{a}_0 e^{\lambda}$. The only difference from the construction developed section \ref{4v}, 
is that in this situation we must apply the operator $a^\dagger$ twice to the state $\psi_a$. Hence, we take the result of equation (\ref{27}) and apply once again the operator 
$a^\dagger$,  
\begin{eqnarray}
a^\dagger \Big(-z B \tilde{a}_0 e^{\lambda}\Big)= (z B)^2 \tilde{a}_0 e^{\lambda}
\label{}
\end{eqnarray}
Following, the same mechanism of the previous section, we should apply the operator $a$ to $(z B)^2 \tilde{a}_0 e^{\lambda}$. This, lead us to,
\begin{eqnarray}
a\Big((z B)^2 \tilde{a}_0 e^{\lambda}\Big) = 4 \tilde{a}_0 B^2 z e^{\lambda} + \tilde{a}_0 
\frac{B^3}{2}[z^2,z^\dagger] e^{\lambda}
\label{}
\end{eqnarray}
The commutator may be calculated easily, being $[z^2,z^\dagger]= \frac{4}{B}z$. Hence, the last equation reduce to  
\begin{eqnarray}
a\Big((z B)^2 \tilde{a}_0 e^{\lambda}\Big) = 6 \tilde{a}_0 B^2 z e^{\lambda}
\label{}
\end{eqnarray}
We need to apply the operator $a$ once again,
\begin{eqnarray}
a\Big( 6 \tilde{a}_0 B^2 z e^{\lambda} \Big) &=& 12 \tilde{a}_0 B^2 e^{\lambda} + 3 \tilde{a}_0 B^3 
[z,z^\dagger] e^{\lambda} 
\nonumber \\
&=& 12 \tilde{a}_0 B^2 e^{\lambda} + 6 \tilde{a}_0 B^2 e^{\lambda} = 18 B^2 (\tilde{a}_0 e^{\lambda})
\label{}
\end{eqnarray}
Then, we have found that $\tilde{a}_0 e^{\lambda}$ is an eigenstate of $a^2 \Big(a^\dagger
\Big)^2$ with eigenvalue $18 B^2$. Therefore, we have  
\begin{eqnarray}
\Big(a^\dagger
\Big)^2 \psi_b = E \psi_a
\label{be1}
\end{eqnarray}
\begin{eqnarray}
a^2 \psi_a = E \psi_b
\label{be2}
\end{eqnarray}
where $E = \pm\sqrt{18 B^2}$, $\chi_a = \frac{(z B)^2}{\pm \sqrt{18B^2}} \tilde{a}_0 
e^{\lambda}$ and $\chi_b = \tilde{a}_0 e^{\lambda}$. Hence, the firts excited enegry eigenstate corresponding to the bilayer 
Hamiltonian (\ref{3dw3}) is 
\begin{eqnarray}
X_{0,1} =\left( \begin{array}{c}
\frac{(z B)^2}{\pm \sqrt{18B^2}} \tilde{a}_0 
e^{\lambda}\\
\tilde{a}_0 e^{\lambda} \end{array} \right)
\label{bilaspin}
\end{eqnarray}
The spinor (\ref{bilaspin}) was obtained from the first of the zero mode in the equation (\ref{zerobi}). However, there is a 
second zero mode, 
\begin{equation}
\left( \begin{array}{c}
z \tilde{a}_0 e^{\lambda}\\
0 \end{array} \right)
\label{xxv}
\end{equation}
We proceed in similar form to previous process. The application of twice the operator $a^\dagger$ upon the upper component of the spinor (\ref{xxv}) give,
\begin{eqnarray}
a^\dagger \Big(z \tilde{a}_0 e^{\lambda} \Big)= \tilde{a}_0 B z^2 e^{\lambda}
\label{}
\end{eqnarray}
\begin{eqnarray}
a^\dagger \Big(\tilde{a}_0 B z^2 e^{\lambda}\Big)= \tilde{a}_0 B^2 z^3 e^{\lambda}
\label{}
\end{eqnarray}
To this result we apply the operator $a$ twice,
\begin{eqnarray}
a \Big(\tilde{a}_0 B^2 z^3 e^{\lambda}\Big) = 6 \tilde{a}_0 B^2 z^2 
e^{\lambda} + \frac{\tilde{a}_0}{2} B^3 [z^3,z^\dagger] e^{\lambda} = 9 \tilde{a}_0 B^2 z^2 e^{\lambda}
\label{}
\end{eqnarray}
\begin{eqnarray}
a \Big(9 \tilde{a}_0 B^2 z^2 e^{\lambda}\Big) = 36 \tilde{a}_0 B^2 
z e^{\lambda} + \frac{9}{2}\tilde{a}_0 B^3 [z^2,z^\dagger]e^{\lambda} = 54 B^2 \Big(z \tilde{a}_0 e^{\lambda}\Big)
\label{1bi}
\end{eqnarray}
Thus, $z \tilde{a}_0 e^{\lambda}$ is an eigenstate of $a^2 \Big(a^\dagger\Big)^2$ with eigenvalue $54 B^2$. This, implies that 
\begin{eqnarray}
X_{0,1}^1 =\left( \begin{array}{c}
\frac{\tilde{a}_0}{\pm\sqrt{54 B^2}}B^2 z^3 e^{\lambda}\\
\tilde{a}_0 z e^{\lambda} \end{array} \right)
\label{80}
\end{eqnarray}
is an eigenstate of the bilayer Hamiltonian with an eigenvalue  $E =\pm\sqrt{54 B^2}$.
We can repeat this method to obtain the higher energy levels and the corresponding eigenstates, finding that the specrum of energies are dictated by the formula,
\begin{eqnarray}
E = \pm\sqrt{9 B^2 N(N-1)},\;\;\;\; N= 0,1,2,3...
\label{nclandau}
\end{eqnarray}
Notice that in the commutative case, the formula for the energy spectrum is (for review see the reference \cite{6}, \cite{mc}, \cite{gc1}), 
\begin{eqnarray}
E_{c} = (2|B|)\pm\sqrt{N(N-1)},\;\;\;\; N= 0,1,2,3...
\label{forcon}
\end{eqnarray}
This last expression coincide with the first non-zero eigenenergy 
calculated for the bilayer graphene in Ref.\cite{6}, \cite{mc}, \cite{gc1}. As in the case of single layer studied in the previous section, this suggest us that energy spectrum is shifted as a result of the space become noncommutative.

\section{Conclusion}
In summary, we have developed a method that allow us to construct the eigenenergies and the eigenstates for Landau levels in 
single  layer and bilayer graphene. In particular, we have been interested in the application of this method to case in which the 
planar coordinates are noncommutative. We have deduced general formulas for the spectrum of enegry, both, for single layer and 
bilayer cases. For the case of single layer we have found that the spectrum of the energy is dictated by the formula,
\begin{eqnarray}
E = \pm 
\sqrt{3NB} \;, \;\;\;
\;\;\; N= 
0,1,2,...
\label{}
\end{eqnarray}
The comparison of this formula with the formula of the spectrum in commutative case, leads us to conclude that the spectrum is 
shifted by $\pm (\sqrt{3} -\sqrt{2})\sqrt{NB}$.  
\\
In addition we have analyzed the bilayer case. In that case we have found that the spectrum of energy is governed by 
\begin{eqnarray}
E = \pm\sqrt{9 B^2 N(N-1)},\;\;\;\; N= 0,1,2,3...
\label{}
\end{eqnarray}
Again, if we compare this formula with that corresponding bilayer graphene in commutative plane, we find that spectrum of energy 
is shifted. The shift, in this case, is $\pm(\sqrt{9}-\sqrt{4})\sqrt{B^2 N(N-1)}$.

\vspace{0.6cm}

{\bf Acknowledgements}
\\
This work is supported by CONICET.

\end{document}